\documentstyle[12pt,epsf]{article}
\begin{document}
\noindent
\begin{flushright}
\vbox{
{\bf TTP93-31}\\
{\bf October 1993}\\
 }
\end{flushright}
\begin{center}
  \begin{large}
COMMENT ON THE AVERAGE MOMENTUM OF TOP QUARKS IN THE THRESHOLD REGION.
\footnote{\normalsize Contributed to the Workshop on Physics at a Linear
Collider, to be published in the proceedings}
\\
  \end{large}

  \vspace{0.8cm}
  \begin{large}
   M. Je\.zabek${}^{a,}\,{}^b$, J.H. K\"uhn${}^b$ and T. Teubner${}^b$ \\
  \end{large}
  \vspace{0.3cm}
${}^a$ Institute of Nuclear Physics, Kawiory 26a, PL-30055 Cracow \\
${}^b$ Institut f\"ur Theoretische Teilchenphysik,Universit\"at Karlsruhe,
    76128 Karlsruhe \\
  \vspace{1.0cm}
  {\bf Abstract}
\end{center}
\noindent
\begin{small}
The behavior of the momentum distribution of top quarks in the threshold
region is investigated. The qualitative behavior, in particular the
dependence of the average momentum on the strong coupling constant
can be understood from analytical calculations for the Coulomb potential.
Ambiguities in the relation between the excitation curve and the top mass
are addressed.
\end{small}
  \vspace{2.0cm}

\noindent 
\newcommand{\pav}{\langle p\rangle}
Of particular interest for top quark studies in the threshold region
is the dependence of the total and the
differential cross section on the strong coupling constant. Some
intuition and qualitative understanding can already be gained from the
predictions based on a pure Coulomb potential.

For a stable top quark of fixed mass the ``effective threshold''
can be associated with the location of the $1S$ resonance
$\sqrt{s_{thr}} = 2 m_t + E_{1S}$ with $E_{1S}=-E_{Ryd}= - \alpha^2m_t/4$
which decreases with increasing $\alpha$. The height of the resonance
cross
section is proportional to the square of the wave function at the origin
and hence proportional to $\alpha^3$, as long as the the resonances are
reasonably well separated. In the limit of large $\Gamma_t$, that is far
larger than $E_{Ryd}$, the overlapping $1S$, $2S$ $\dots$ resonances have
to fill the gaps between the peaks. Since these gaps themselves
increase proportional to $\alpha^2$, one is left in the extreme case of large
width with a cross section linear in $\alpha$. Note that this
corresponds to the behavior of the cross section close to but slightly
above the threshold which is also proportional to $\alpha$.

For realistic top masses  of about 150 GeV one thus observes a behavior
of the peak cross section quite close to the first power in $\alpha$.
Since the location of the peak itself depends on $\alpha$, only the
analysis of the full shape allows to extract the relevant information.

Once the threshold energy is determined experimentally, $\alpha$ and
$m_t$ are still strongly correlated and can hardly be deduced
individually.

Therefore in a next step also the momentum distribution of top quarks
has to be
exploited to obtain further information. The discussion is again
particularly simple for the Coulomb potential $V(r)=-\alpha/r$.
The average momentum,
in units of the Bohr momentum $\alpha m_t/2$, can be written in terms of
a function $f(\epsilon)$ which depends only on one variable $\epsilon=
E/E_{Ryd}$ if
the energy $E=\sqrt{s}- 2m_t$ is measured in terms of the Rydberg
energy.

\begin{equation}\label{paveq}
\pav = \frac{\alpha m_t}{2} f(\epsilon)
\end{equation}
For positive arguments  the function $f$ can be derived
from obvious kinematical considerations.

\begin{equation}
f(\epsilon)=\sqrt{\epsilon} \qquad for \qquad \epsilon \ge0
\end{equation}

For the discrete negative
arguments $\epsilon_n=-1/n^2$ corresponding to the
locations of the bound states
the radial wave functions in momentum space are given \cite{Bethe}
in terms  of the Gegenbauer polynomials $C^m_n$

\begin{equation}
\psi(\vec p\,)= \frac{16\pi n^{3/2}}{(1+n^2 p^2)^2} C^1_{n-1}
\left(\frac{n^2 p^2 -1} {n^2 p^2 + 1} \right) Y_0^0\left(\theta,\varphi\right)
\end{equation}
with
\begin{equation}
\int\frac {d\vec p}{(2\pi)^3} |\psi(\vec p\,)|^2 = 1\ .
\end{equation}
Using the explicit forms of $C^m_n$
\begin{equation}
C^1_0(z)=1,\qquad C^1_1(z)=2z,\qquad C^1_2(z)=4z^2-1
\end{equation}
one obtains through straightforward calculation

\begin{equation}
f(-1)=\frac{8}{3\pi},\qquad f(-1/4)=\frac{16}{15\pi},
\qquad f(-1/9)=\frac{24}{35\pi}\ .
\end{equation}
For arbitrary $n$ one may exploit the following identity
\begin{equation}
C^1_n(\cos\phi)=\frac{\sin (n+1)\phi}{\sin\phi}
\end{equation}
to derive the general result
\begin{equation}
f\left(-\frac{1}{n^2}\right)=\frac{8n}{(2n-1)(2n+1)\pi}
\label{eq8}
\end{equation}
with the asymptotic behavior
\begin{equation}\label{asym}
f\left(-\frac{1}{n^2}\right)\to \frac{2}{n \pi}\ .
\end{equation}

This in accord with the result expected from classical mechanics:
For the average momentum of a particle on a closed orbit
in the Coulomb potential one derives
\begin{equation}
\langle p^{2n}\rangle=\left(\frac{\alpha m_t}{2}\right)^{2n}
\left(\frac{-E}{E_{Ryd}}\right)^n
\frac{1}{2\pi}
\int_0^{2\pi} d\xi
\frac{(1-e^2\cos^2\xi)^n}{(1-e\cos\xi)^{2n-1} }\ .
\end{equation}
Quantum mechanical orbits with angular momentum zero and
high radial quantum numbers correspond to classical motions
with excentricity $e=1$. In this limiting case the classical
expectation value is easily evaluated and for $n=1/2$ one finds
agreement with the quantum mechanical result.
For small negative energies one
therefore obtains the behavior $f(\epsilon) = 2\sqrt{-\epsilon}/\pi$.
Significantly below threshold, however, the average momentum increases
more rapidly with decreasing energy and between the $1S$ and the $2S$
state one observes an approximately
linear dependence on the energy.

From these considerations the dependence of the average momentum
on $\alpha$ (with $E$ fixed) is easily understandable, in particular
the seemingly surprising observation that well below threshold
$\pav$ decreases with
increasing $\alpha$.
From (\ref{paveq}) one derives for
a shift in $\alpha$ (keeping the energy $E$ fixed) the following
shift in $\pav$
\begin{equation}
\frac{\delta\pav}{\pav}=
\left(1-2\frac{f'(\epsilon)}{f(\epsilon)} \epsilon\right)
\frac{\delta\alpha}{\alpha}\ .
\end{equation}
Above threshold as well as close to but below threshold $f\propto
\sqrt{|\epsilon|}$.
Hence $\epsilon f'/f = 1/2$ and the average momentum remains unaffected.
Significantly below threshold, however, $\epsilon f'/f \approx 1$
and the factor in front of $\delta\alpha/\alpha$ becomes negative.
This explains why $\pav$ decreases with increasing $\alpha$.

\begin{figure}
\epsffile{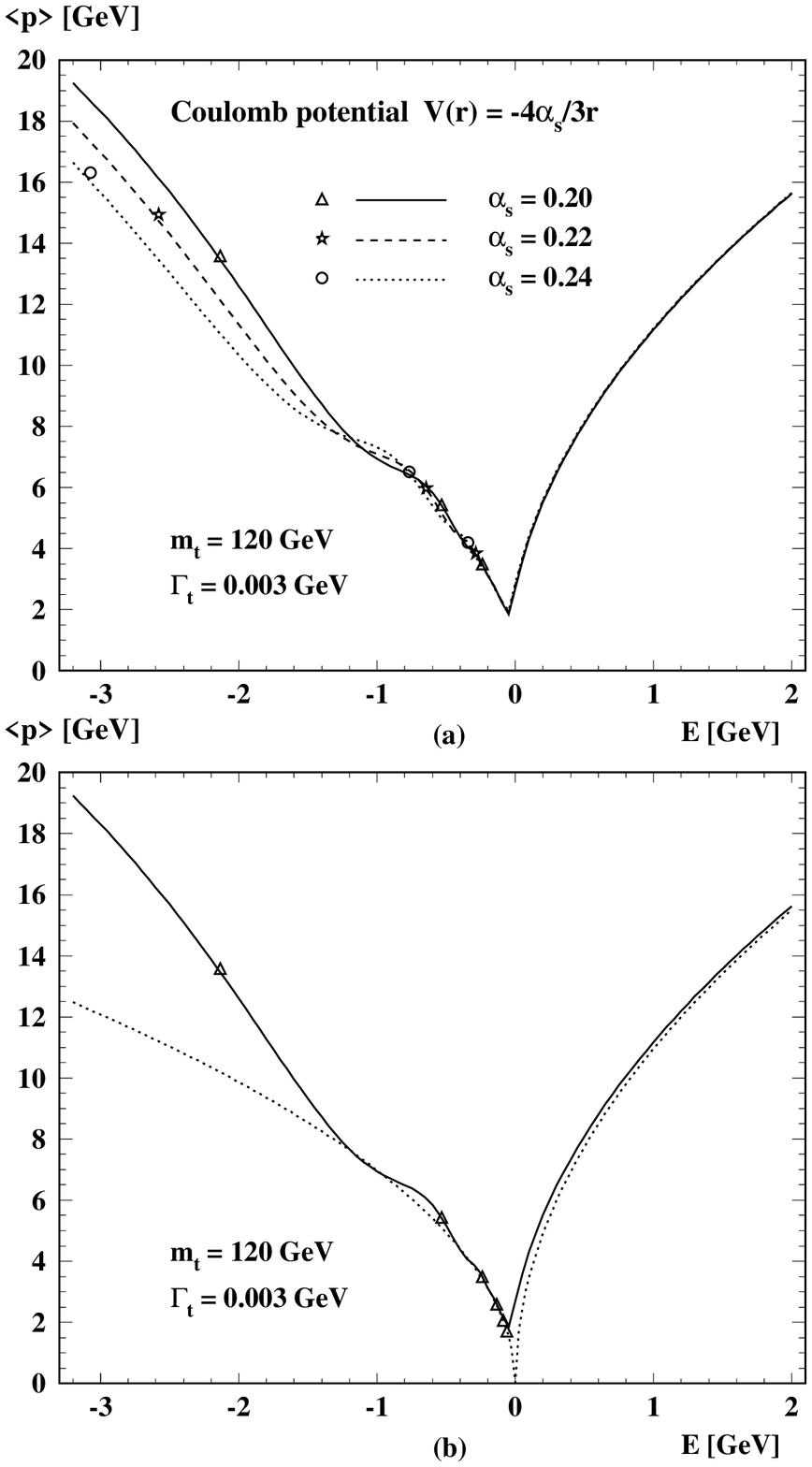}
\caption{a) Average momentum as a function of $E$ for different values
of $\alpha$. The markers show the results of the analytical
calculation at $1S$, $2S$, $3S$ energies.
b) Comparison with the analytical result for discrete
energies and with the square--root dependence close to threshold.}
\label{fig1}
\end{figure}

These results are illustrated in Fig.\ref{fig1}.
In Fig.\ref{fig1}b we demonstrate that $\pav$ as evaluated with the
program for the Green function (solid line) coincides perfectly
well with the values calculated from the analytical formula on
resonance, indicated
by the triangles. The prediction from classical mechanics, namely
$\pav \propto \sqrt{|\epsilon|}$ is shown by the dotted line
and agrees nicely for positive and negative energies. In Fig.\ref{fig1}a
$\alpha_s$ is increased from 0.20 to 0.24 and $\pav$ changes in accord
with the previous discussion.

For definiteness we have
chosen $m=m_t/2=60$ GeV for the reduced mass and
$\alpha=4\alpha_s/3$ with $\alpha_s$ varying between 0.20 and 0.24.
To retain the stability of our numerical program $\Gamma_t=3$ MeV has
been kept non vanishing and a cut $p<m_t/2$ has been introduced.
The curves demonstrate the increase of $\pav$ by
about 10\%  for the corresponding increase in $\alpha$. The triangles
mark the locations of the resonances and the expectation values
for the momentum as calculated from (\ref{eq8}).

\begin{figure}
\epsffile{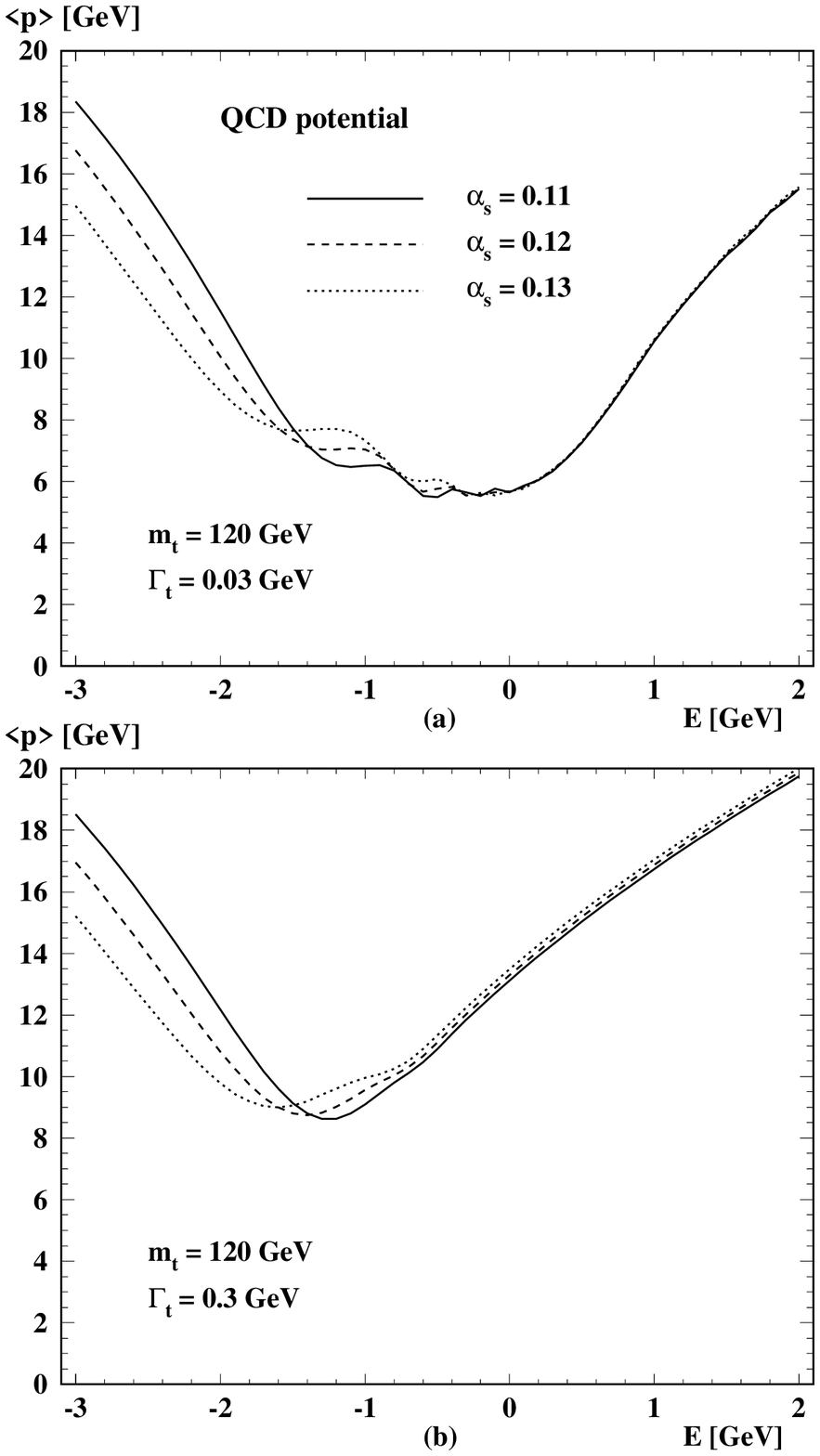}
\caption{Energy dependence of $\langle p\rangle$, the
average t quark momentum
for $\alpha_s = 0.13$ (dotted)
$0.12$ (dashed) and $0.11$ (solid) line for
$m_t=120$ GeV. a) $\Gamma_t=0.03$ GeV and b) $\Gamma_t=0.3$ GeV.}
\label{fig2}
\end{figure}

The qualitative behavior remains unchanged for realistic QCD
potentials.
Predictions for different QCD potentials based on the techniques as
described in \cite{TJ,JKT} are shown in Fig.\ref{fig2}.
These potentials correspond to different values of $\alpha_s(M_Z)$.
Qualitatively the same behavior is observed as in Fig.\ref{fig1}.
In Fig.\ref{fig2}a the top quark width has been set to an artificially
small value of $0.03$ GeV, in Fig.\ref{fig2}b the realistic value of
$0.3$ GeV has been adopted.

An important feature is evident from Fig.\ref{fig2}: The momentum
calculated for
positive energy is nearly independent from $\alpha_s$ and reflects
merely the kinematic behavior, just as in the case of the Coulomb
potential. This is characteristic for the choice of a potential \cite{TJ}
where the large distance behavior is fixed by phenomenology and
decoupled from the short distance value of $\alpha_s$.

At this point a brief discussion is in order on the relation
between the top masses determined on the basis of different
potentials.  The perturbative two loop QCD
potential has been calculated in momentum space and is fixed
unambiguously for sufficiently large $Q^2$:
\begin{equation}
V\left( Q^2 , \alpha _{\overline{MS}} (Q^2)\right) = - \frac{16\pi}{3}
 \frac{\alpha _{\overline{MS}} (Q^2)}{Q^2}
  \left[ 1 + \left( \frac{31}{3} -
  \frac{10}{9} n_f \right)
   \frac{\alpha _{\overline{MS}}(Q^2)}{4\pi} \right]
\label{eq:49}
\end{equation}
with
\begin{eqnarray}
\frac{\alpha_{\overline{MS}}(Q^2)}{4\pi } &=&
\frac{1}{b_0 \log
 \left( Q^2/\Lambda^2_{\overline{MS}}
 \right)}
 \left[    1 - \frac{b_1}{b^2_0}
          \frac   {\log\log
          \left(  Q^2 / \Lambda^2_{\overline{MS}} \right) }
                  {\log\left( Q^2 / \Lambda^2_{\overline{MS}}
          \right) }
 \right]\quad
\label{eq:50}\\
b_0 &=& 11 - \frac{2}{3}n_f, \qquad
b_1 = 102 - \frac{38}{3}n_f
\nonumber
\end{eqnarray}
Power law suppressed terms
cannot be calculated in this approach.  To evaluate the Fourier
transform of this function in order to calculate the potential in
coordinate space the small $Q^2$ behavior has to be specified in
an ad hoc manner.  Different assumptions will lead to the same
short distance behavior.  The potentials will, however, differ
with respect to their long distance behavior.  In \cite{khoze} it
has been argued convincingly that the long distance tail is cut off
by the large top width.  However, an additive constant in coordinate
space can be induced by the small momentum part of $\tilde V(Q^2)$.
This additional term would lead to a shift in the $t\bar t$ threshold,
which in turn can be reabsorbed by a corresponding shift in $m_t$.
The different assumptions on the long distance behavior are reflected
in differences between
the predictions of \cite{peskin,Sumino,JKT}
for the precise location of the $t\bar t$
threshold for identical values of $\alpha_s$ and $m_t$ and in
differences
in the $\alpha_s$ dependence of the momentum distributions for fixed
$m_t$ and energy (see also \cite{Martinez}).
All these differences can be attributed to the
freedom in the additive constant discussed before.  The same additive
constant appears in $b\bar b$ spectroscopy, such that the mass
difference between top and bottom is independent from these
considerations.

\sloppy
\raggedright
\def\app#1#2#3{{\it Act. Phys. Pol. }{\bf B #1} (#2) #3}
\def\apa#1#2#3{{\it Act. Phys. Austr.}{\bf #1} (#2) #3}
\def\lhc{Proc. LHC Workshop, CERN 90-10}
\def\npb#1#2#3{{\it Nucl. Phys. }{\bf B #1} (#2) #3}
\def\plb#1#2#3{{\it Phys. Lett. }{\bf B #1} (#2) #3}
\def\prd#1#2#3{{\it Phys. Rev. }{\bf D #1} (#2) #3}
\def\pR#1#2#3{{\it Phys. Rev. }{\bf #1} (#2) #3}
\def\prl#1#2#3{{\it Phys. Rev. Lett. }{\bf #1} (#2) #3}
\def\prc#1#2#3{{\it Phys. Reports }{\bf #1} (#2) #3}
\def\cpc#1#2#3{{\it Comp. Phys. Commun. }{\bf #1} (#2) #3}
\def\nim#1#2#3{{\it Nucl. Inst. Meth. }{\bf #1} (#2) #3}
\def\pr#1#2#3{{\it Phys. Reports }{\bf #1} (#2) #3}
\def\sovnp#1#2#3{{\it Sov. J. Nucl. Phys. }{\bf #1} (#2) #3}
\def\jl#1#2#3{{\it JETP Lett. }{\bf #1} (#2) #3}
\def\jet#1#2#3{{\it JETP Lett. }{\bf #1} (#2) #3}
\def\zpc#1#2#3{{\it Z. Phys. }{\bf C #1} (#2) #3}
\def\ptp#1#2#3{{\it Prog.~Theor.~Phys.~}{\bf #1} (#2) #3}
\def\nca#1#2#3{{\it Nouvo~Cim.~}{\bf #1A} (#2) #3}

%

\begin{thebibliography}{99}
\bibitem{Bethe} H.A.~Bethe and E.E.~Salpeter, {\em Quantum Mechanics
 of One-- and Two--Electron Atoms} (Plenum Publishing Corporation,
 New York, 1977)
\bibitem{TJ} M.~Je\.zabek and T.~Teubner, \zpc{59}{1993}{669}
\bibitem{khoze} V.S.~Fadin, V.A.~Khoze, \jet{46}{1987}{525},
   \sovnp{48}{1988}{309}
\bibitem{peskin}  J.M.~Strassler and M.E.~Peskin, \prd{43}{1991}{1500}
\bibitem{Sumino} Y.~Sumino, K.~Fujii, K.~Hagiwara, H.~Murayama, C.-K.~Ng,
   \prd{47}{1992}{56}
\bibitem{JKT}  M.~Je\.zabek, J.H.~K\"uhn and T.~Teubner,
   \zpc{56}{1992}{653}
\bibitem{Martinez} P.~Igo--Kemenes, M.~Martinez, R.~Miquel, S.~Orteu,
contribution to this workshop
\end{thebibliography}
\end{document}